\documentstyle[aps,subfigure,epsfig,prl,amssymb,multicol,citesort]{revtex}
\def\unitr{{\mbox{\boldmath$\hat r$}}}

\def\unitz{{\mbox{\boldmath$\hat z$}}} 
\def\f{{\mbox{\boldmath$f$}}}

\def\k{{\mbox{\boldmath$k$}}}

\def\v{{\mbox{\boldmath$v$}}}

\newcommand{\x}{{\mbox{\boldmath$x$}}}

\def\unitr{{\mbox{\boldmath$\hat r$}}}
\def\unitz{{\mbox{\boldmath$\hat z$}}} 
 
\def\cP{{\cal P}}

\def\dv{{\delta_r v}}
\def\nab{{\bf \nabla}}

\newcommand{\be}{\begin{equation}} 
\newcommand{\ee}{\end{equation}}
\newcommand{\bea}{\begin{eqnarray}} 
\newcommand{\eea}{\end{eqnarray}}
\newcommand{\r}{{\bf r}} 
\newcommand{\lp}{\left(}
\newcommand{\rp}{\right)} 
\newcommand{\la}{\left\langle}
\newcommand{\ra}{\right\rangle}
\begin{document} 
\title{The decay of homogeneous anisotropic turbulence}
\author{L.~Biferale$^{1}$, G. Boffetta$^{2}$, A.~Celani$^{3}$, A.~Lanotte$^{4}$, 
F.~Toschi$^{5}$, and M. Vergassola$^{6}$} 
\address{$^1$ Dipartimento di Fisica, Universit\`a "Tor Vergata", and INFM, Unit\`a di Tor Vergata,\\ Via della Ricerca Scientifica 1, I-00133 Roma, Italy} 
\address{$^2$ Dipartimento di Fisica Generale, Universit\`a di Torino,\\
 Via Giuria 1, I-10125, Torino, Italy and INFM, Sezione di Torino Universit\`a} 
 \address{$^3$ CNRS, INLN, 1361 Route des Lucioles, F-06560 Valbonne, France}  
 \address{$^4$ CNR, ISAC - Sezione di Lecce, Str. Prov. Lecce-Monteroni km. 1200, I-73100 Lecce,
 Italy and \\INFM, Unit\`a di Tor Vergata, I-00133 Roma, Italy}
 \address{$^5$ Istituto per le Applicazioni del Calcolo, CNR, Viale del Policlinico 137, 
I-00161 Roma,\\ and INFM, Unit\`a di Tor Vergata, I-00133 Roma, Italy} 
 \address{$^6$ CNRS, Observatoire de la C\^{o}te d'Azur, B.P. 4229, F-06304 Nice Cedex 4, France}

\maketitle
\begin{abstract}
We present the results of a numerical investigation of
three-dimensional decaying turbulence with statistically homogeneous
and anisotropic initial conditions. We show that at large times, in
the inertial range of scales: (i) isotropic velocity fluctuations
decay self-similarly at an algebraic rate which can be obtained
by dimensional arguments; (ii) the ratio of anisotropic to isotropic
fluctuations of a given intensity falls off in time as a power law,
with an exponent approximately independent of the strength of the
fluctuation; (iii) the decay of anisotropic fluctuations is not
self-similar, their statistics becoming more and more intermittent as
time elapses. We also investigate the early stages of the decay.
The different short-time behavior observed in two experiments differing 
by the phase organization of their initial conditions gives a new hunch on 
the degree of universality of small-scale turbulence statistics, i.e. its
independence of the conditions at large scales.
\end{abstract}
%%%%%%%%%%%%%%%%%%%%%%% 
\begin{multicols}{2}
\section{Introduction} 
Decaying turbulence has attracted the attention of various
communities and is often considered in experimental, numerical and
theoretical investigations \cite{batche,frisch,my}. It is in fact
quite common that even experiments aimed at studying stationary
properties of turbulence involve processes of decay. Important
examples are provided by a turbulent flow behind a grid (see
\cite{stelp99} and references therein) or the turbulent flow created
at the sudden stop of a grid periodically oscillating within a bounded
box \cite{desilva}.  In the former case, turbulence is slowly decaying
going farther and farther away from the grid and its characteristic
scale becomes larger and larger (see \cite{stelp99} for a thorough
experimental investigation). Whenever there is sufficient separation
between the grid-size $L_{in}$ and the scale of the tunnel or the tank
$L_{0} \gg L_{in}$, a series of interesting phenomenological
predictions can be derived.  For example, the decay of the
two-point velocity correlation function, for both isotropic and
anisotropic flows, can be obtained under the so-called
self-preservation hypothesis (see \cite{my} chapter XVI). That posits
the existence of rescaling functions allowing to relate correlation
functions at different spatial and temporal scales. By inserting this
assumption into the equations of motion, asymptotic results can be
obtained both for the final viscosity-dominated regime and for the
intermediate asymptotics when nonlinear effects still play an
important role. \\
The status of the self-preservation hypothesis and the
properties of energy decay in unbounded flows are still controversial
\cite{frisch,my,dec2,stelp99}. Systematic results on related problems
have been established recently, e.g. for non-linear models of
Navier-Stokes equations as Burgers' equation, see e.g. \cite{burgers_decay}, 
and for stochastic models of linear passive advection \cite{rev}, both in
unbounded \cite{Son,BF,ex,cefv} and bounded domains \cite{cl,SP}. \\ Here, we
investigate the decay of three-dimensional homogeneous and anisotropic
turbulence by direct numerical simulations of the Navier-Stokes
equations in a periodic box.  Previous numerical studies have been
limited to either homogeneous and isotropic turbulence \cite{bo95,iop}
or to shell models of the energy cascade \cite{dt}.\\  The initial
conditions are taken from the stationary ensemble of a turbulent flow
forced by a strongly anisotropic input \cite{bt00}.  The correlation
lengthscale of the initial velocity field $L_{in}$ is of the order of
the size of the box $L_{0}\approx L_{in}$.\\ In the first part of
this paper, we shall try to answer the following questions about the
intermediate asymptotic regime of nonlinear decay: How do global
quantities, such as single-point velocity and vorticity correlations,
decay~?  What is the effect of the outer boundary on the decay laws~?
Do those quantities keep track of the initial anisotropy~?  As for the
statistics of velocity differences within the inertial range of
scales, is there a recovery of isotropy at large times~?  If so, do
strong fluctuations get isotropic at a faster/slower rate with
respect to those of average intensity~?  Do isotropic and anisotropic
fluctuations decay self-similarly~?  If not, do strong fluctuations
decay slower or faster than typical ones~? \\ In the second part we
study the early stages of the decay, with the aim of establishing a link
between the small-scale velocity statistics in this phase and in the
forced case. That will allow us to argue in favor of an ``exponents
only'' universality scenario, for forced hydrodynamic turbulence.
%%%%%%%%%%%%%%%%%%%%%%%%%%%%%%%%%%%%%%%%%%%%%%%%%%%%%%
\section{Numerical setup}
\label{details}
\subsection{The initial conditions} 
The initial conditions are taken from the stationary ensemble of a
forced random Kolmogorov flow \cite{bt00}. For sake of completeness,
we recall here some of the statistical properties of this forced
turbulent flow.  We consider the solutions of the Navier-Stokes
equations for an incompressible velocity field ${\v}$.
\be
\partial_t {\v} + ({\v}\cdot \nab) {\v} = -\nab p + \nu \Delta {\v},
\label{eq:navierstokes}
\ee
in a three-dimensional periodic domain. To maintain a statistically 
stationary state Eq.~(\ref{eq:navierstokes}) had to be supplemented by an 
input term ${\f}$ acting at large scales. This force was strongly anisotropic:
${\f}=(0,0,f_z(x))$ with $f_z(x)=F_1 \cos[2\pi x/L_x +\phi_1(t)] + 
F_2 \cos[4 \pi x/ L_x +\phi_2(t)]$, constant amplitudes $F_{1,2}$ and
independent, uniformly distributed, $\delta$-correlated in time random
phases $\phi_{1,2}(t)$.  This choice ensured the statistical
homogeneity of the forcing and thus of the velocity field.  We
simulated the forced random Kolmogorov flow at resolution $256^3$ for
time spans up to $70$ eddy turnover times \cite{bt00}. The viscous
term was replaced by a second-order hyperviscous term $-\nu \Delta^2{\v}$.  
We stored $40$ statistically independent configurations that
here serve as initial conditions for the decaying runs.
%--------------------------------------------------------
\begin{figure}[hb] \centering
\subfigure[$v_z(\tau_0)$]
{\includegraphics[draft=false,width=2.65cm]{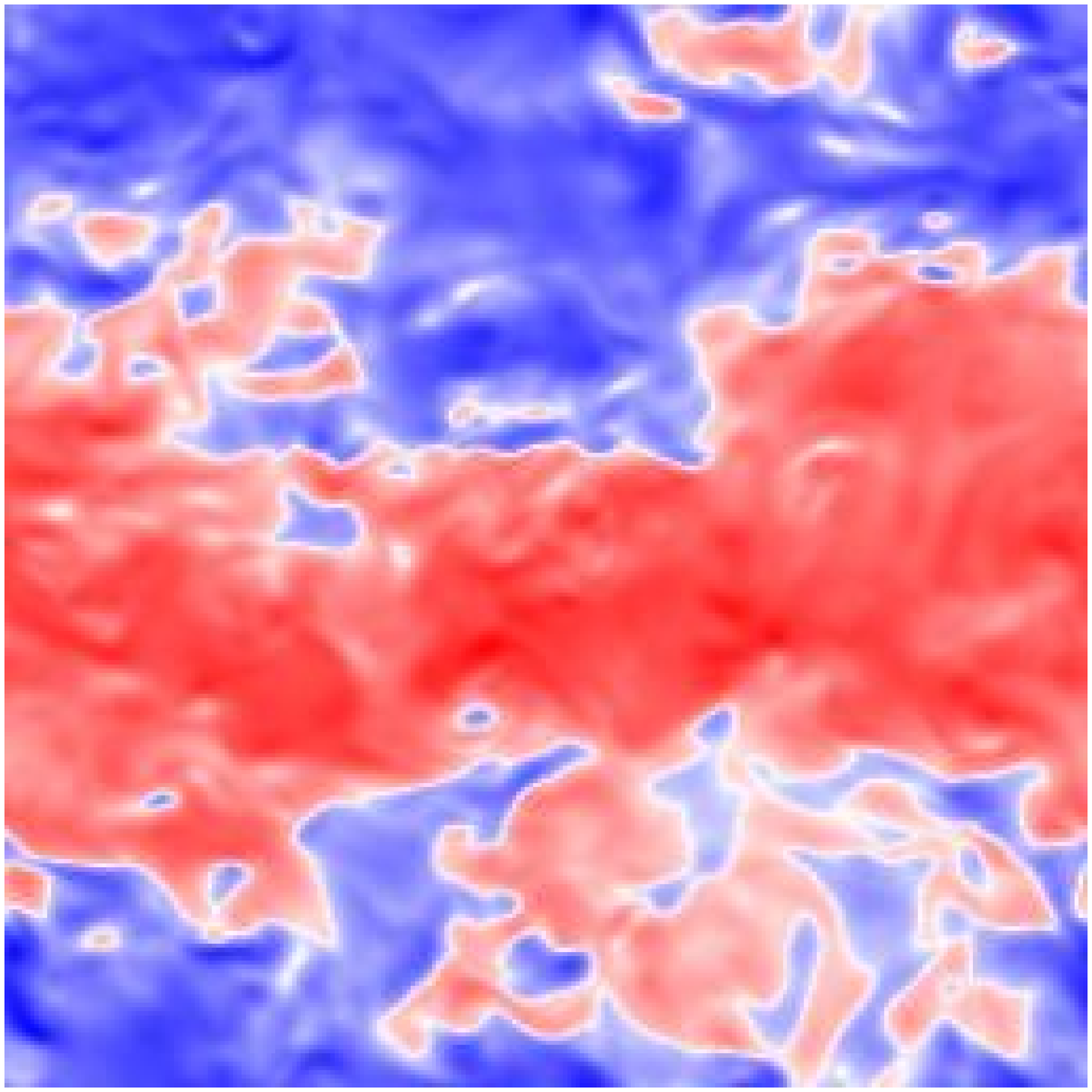}} \hspace{1mm}
\subfigure[$v_z(10\tau_0)$]
{\includegraphics[draft=false,width=2.65cm]{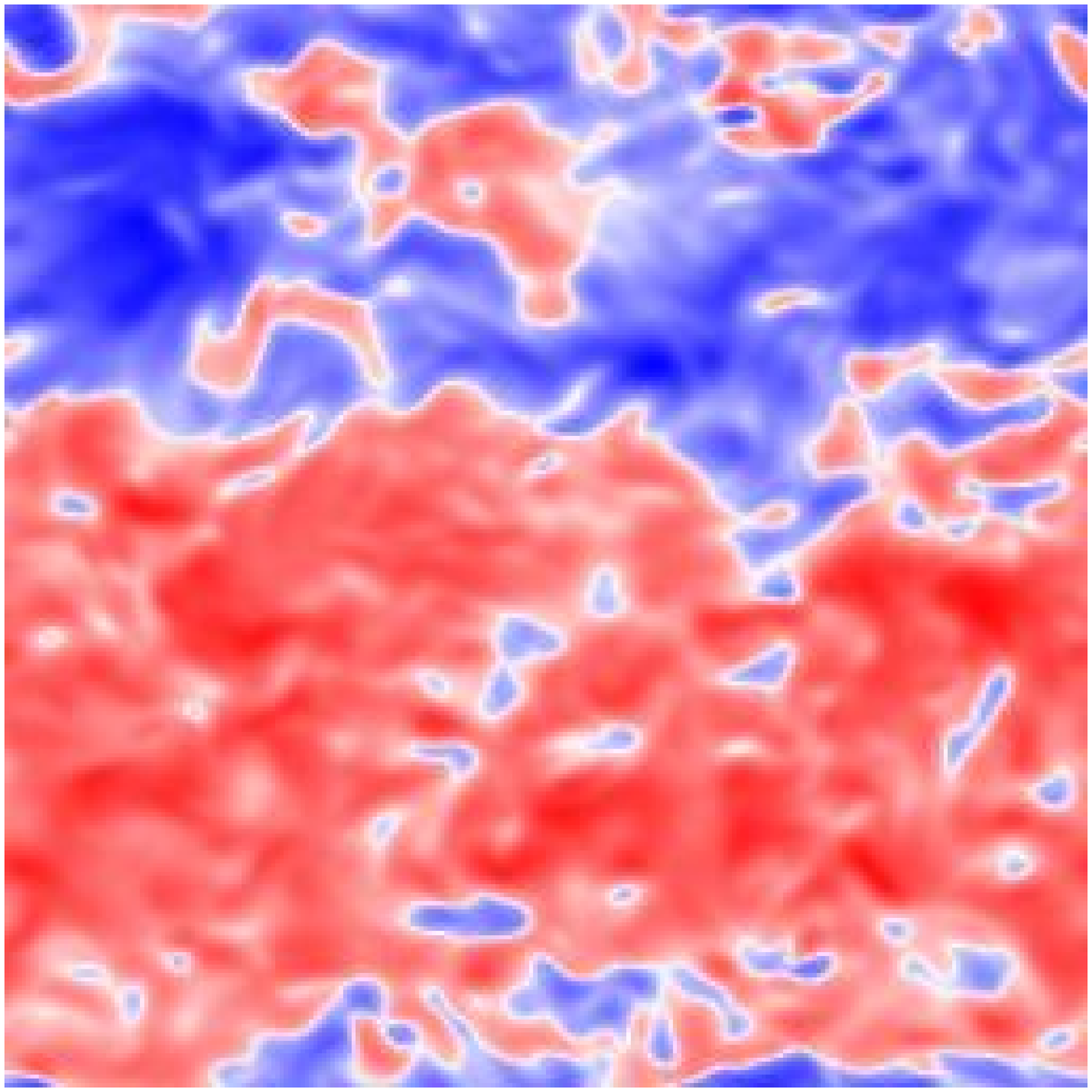}} \hspace{1mm}
\subfigure[$v_z(100\tau_0)$]
{\includegraphics[draft=false,width=2.65cm]{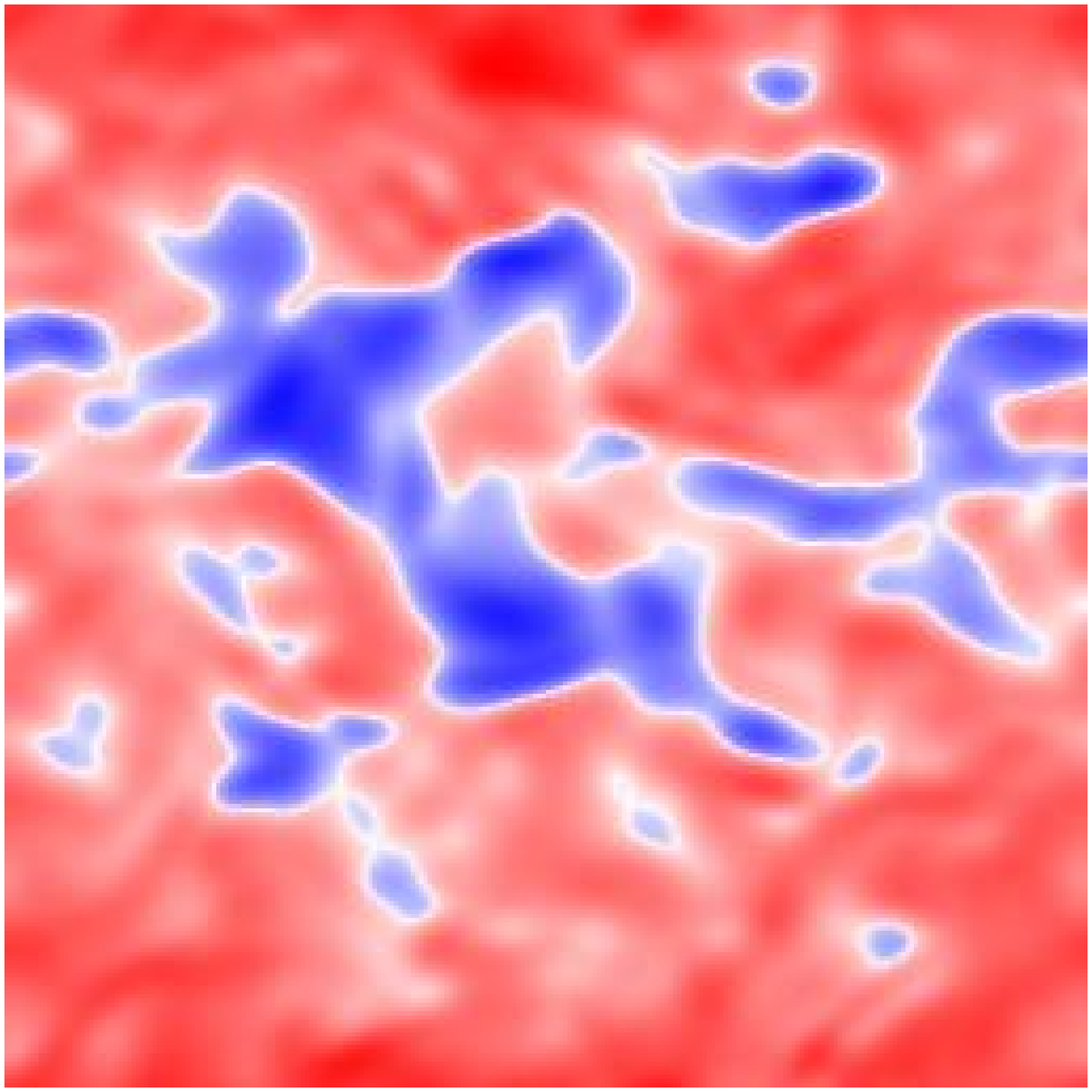}} \hspace{1mm}
\subfigure[$v_x(\tau_0)$]
{\includegraphics[draft=false,width=2.65cm]{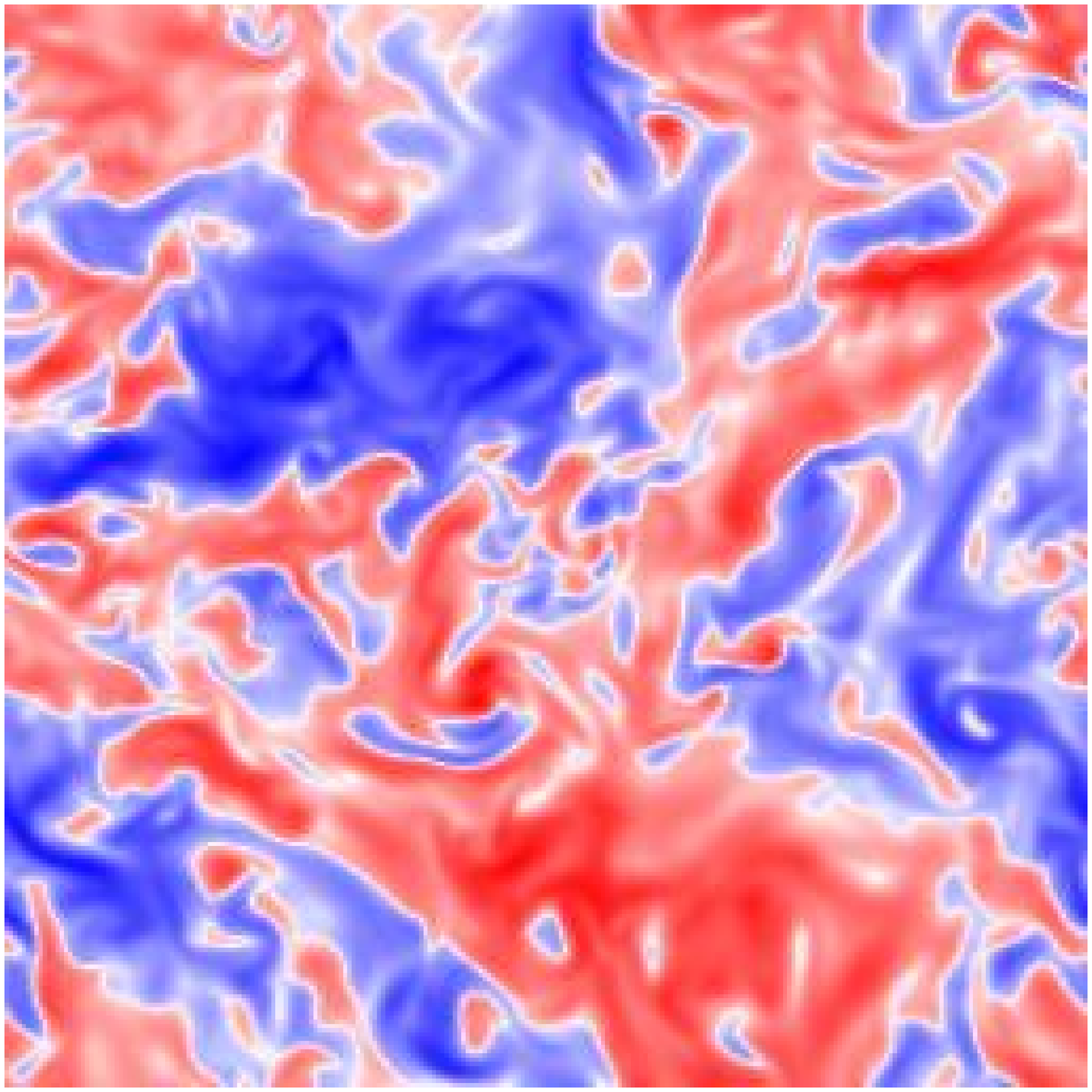}} \hspace{1mm}
\subfigure[$v_x(10\tau_0)$]
{\includegraphics[draft=false,width=2.65cm]{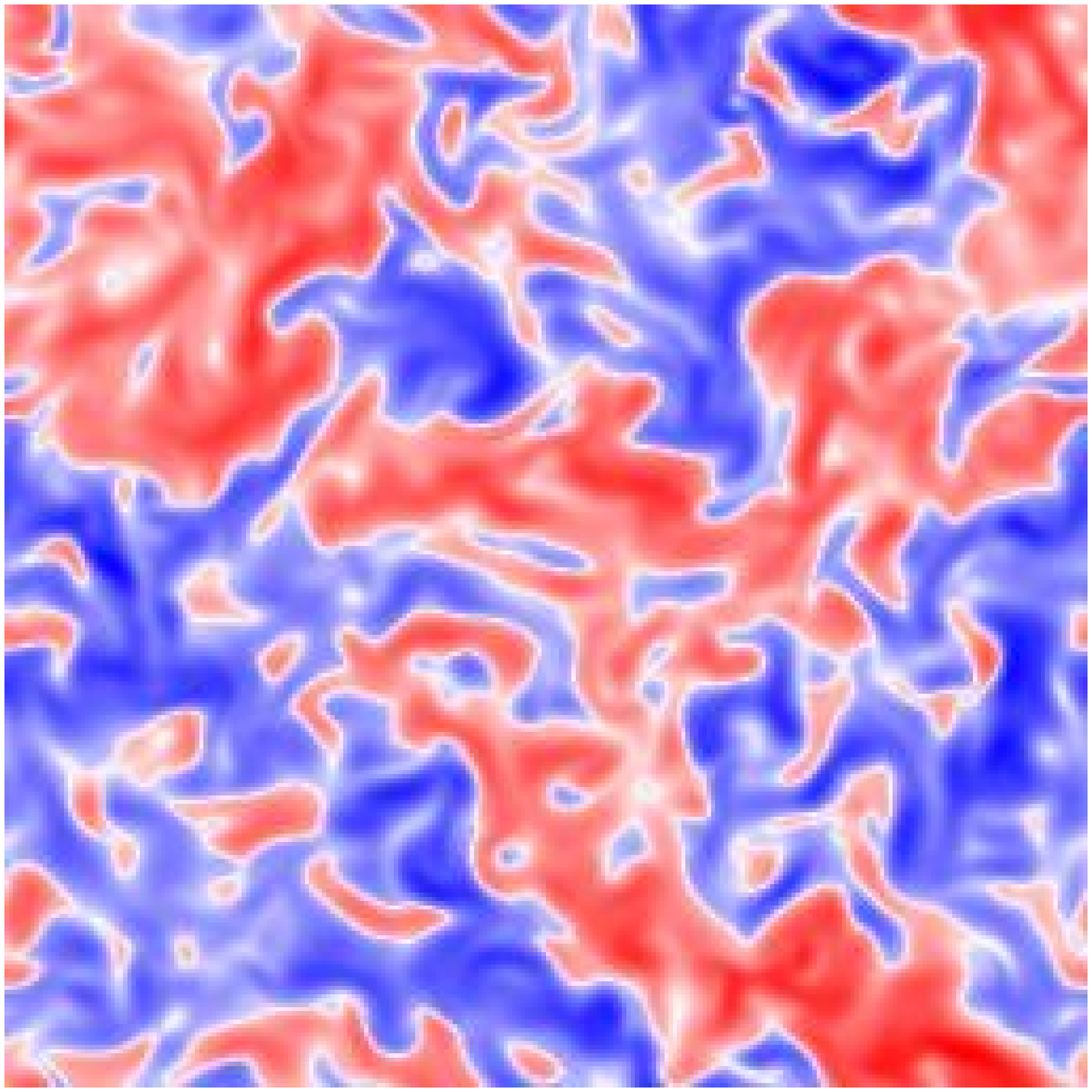}} \hspace{1mm}
\subfigure[$v_x(100\tau_0)$]
{\includegraphics[draft=false,width=2.65cm]{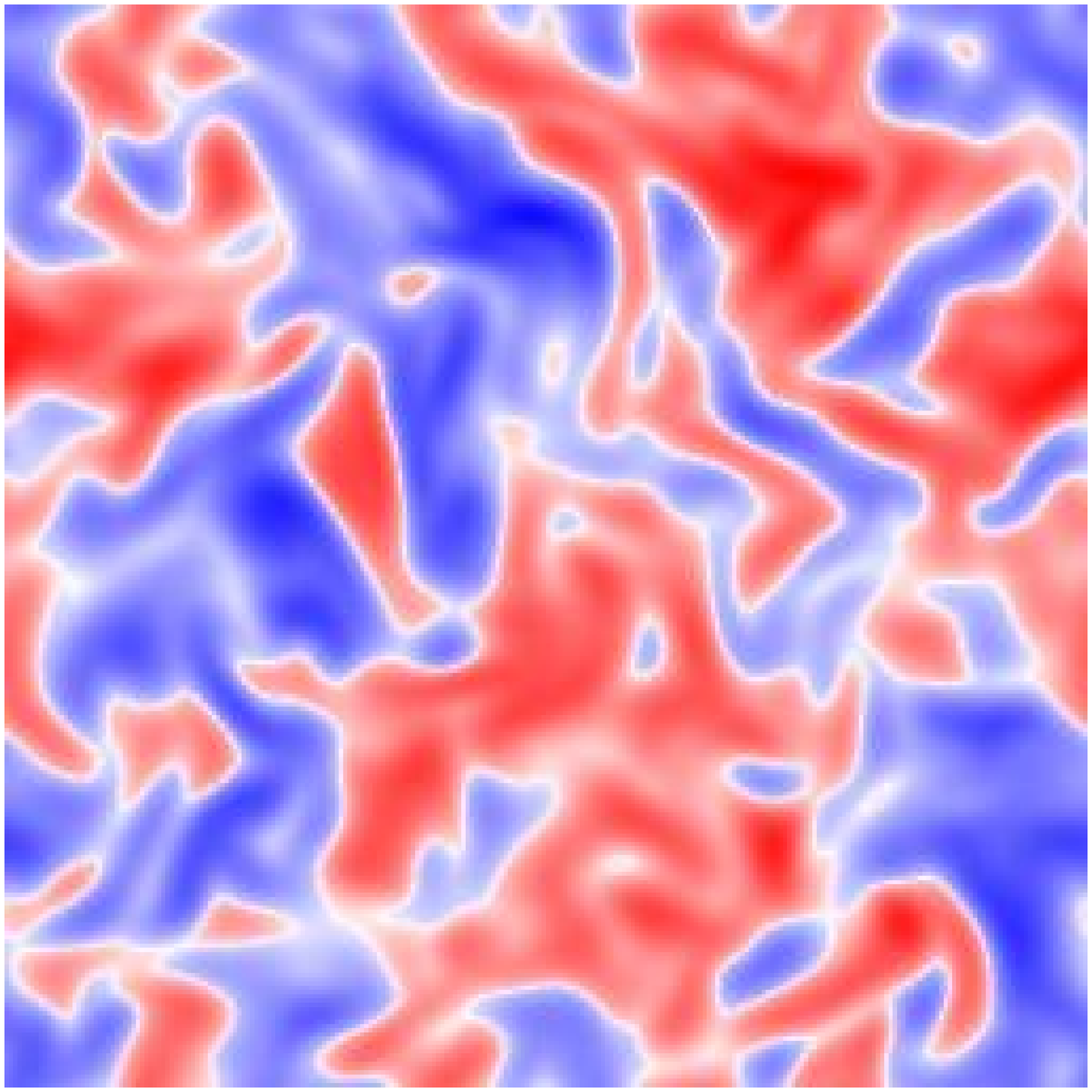}} 
\caption{Two-dimensional sections of a typical velocity field at different
  times of decay. The three top images and the three bottoms are the grayscale
  plots of the velocity components parallel and 
transverse to the direction of the forcing, respectively. Note in the upper
row, the presence of anisotropic structures which decay as time elapses.}
\label{fig:1}
\end{figure}
%--------------------------------------------------------------------
\subsection{Decaying runs}
As turbulence decays, the effective Reynolds number $Re= v_{rms} L_{0} /\nu$ 
decreases, while the viscous characteristic scale and time increase. 
To speed up the numerical time-marching, it is then convenient to use an 
adaptative scheme. We calculate periodically the
smallest eddy-turnover time from the energy spectrum and set the time
step as $1/100$ thereof. The whole velocity-field configuration is then
dumped for offline analysis at fixed multiples
$\{0,1,10,10^{2},10^{3},10^4,10^5,10^6\}\,\tau_0$ of the initial large-scale
eddy turnover time $\tau_0 = L_{0}/v_{rms}^{t=0}$.  In
Fig.~\ref{fig:1} we show a two dimensional section in the plane
$x$-$z$ of the velocity components $v_z$ and $v_x$.
\section{The decay of global quantities}
\begin{figure}[h] 
\epsfxsize=8.0truecm
\epsfysize=6.0 cm
\epsfbox{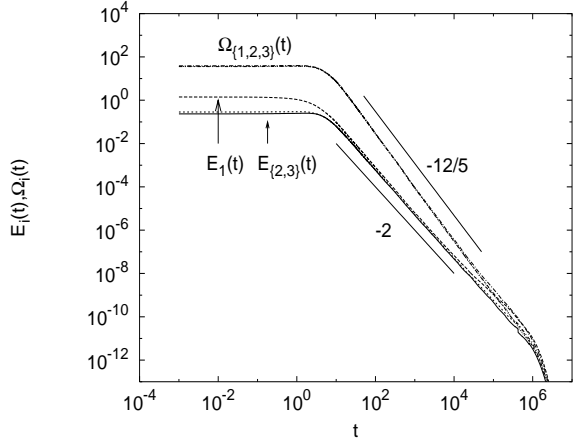} 
\vspace{4pt}
\caption{Log-log plot of the eigenvalues of energy and vorticity matrices 
{\it vs.} time, expressed in $\tau_0$ unit.}
\label{fig:2} 
\end{figure}

\begin{figure}[h] 
\epsfxsize=8.0truecm
\epsfysize=6.0 cm
\epsfbox{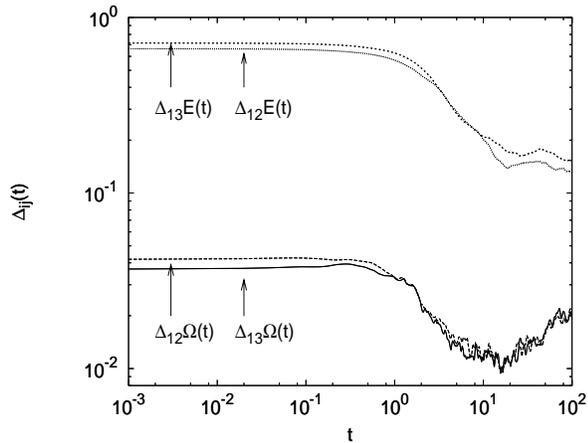} 
\caption{Log-log plot of the anisotropy content at the large scales 
($\Delta_{12}E(t)$, $\Delta_{13}E(t)$, top curves) and the small scales 
($\Delta_{12}\Omega(t)$, $\Delta_{13}\Omega(t)$, bottom curves) as a function
of time, expressed in $\tau_0$ unit. The large-scale (small-scale) anisotropy 
content is defined as the mismatch between the eigenvalues of the single-point 
velocity (vorticity) correlation.}
\label{fig:3} 
\end{figure}

A first hint on the restoration of isotropy at large times can
be obtained by the two-dimensional snapshots in
Fig.~\ref{fig:1}.  After a few eddy turnover times, it is evident that
large-scale fluctuations become more and more isotropic. To give a
quantitative measure, we collect for each run the temporal behavior of
the following one-point quantities: \bea E_{ij} = {\overline{v_i(t)v_j(t)}},\\ 
\Omega_{ij} = {\overline{\omega_i(t) \omega_j(t)}}.  \eea
By $\overline{\cdots}$ we denote the average over space coordinates
only, whereas $\la \cdots \ra$ will indicate the average over both initial
conditions and space. The symmetric matrices $E_{ij}(t)$ and $\Omega_{ij}(t)$
are then diagonalized at each time-step and the eigenvalues
$E_1(t),E_2(t),E_3(t)$ and $\Omega_1(t),\Omega_2(t),\Omega_3(t)$ are
extracted. Since the forcing points in a fixed direction two eigenvalues are
almost degenerate, say $E_2$ and $E_3$, and strongly differ from the first
one, $E_1$.  
The typical decay of $E_{i}(t)$ and $\Omega_{i}(t)$ for $i=1,\dots,3$ is shown
in Fig.~\ref{fig:2}. During the self-similar stage,
$t\in [10,10^6]$, the energy eigenvalues fall off as
$E_{\{1,2,3\}}\sim t^{-2}$, as expected for the decay in a bounded
domain \cite{stelp99,bo95}. The enstrophy eigenvalues,
$\Omega_{\{1,2,3\}}$ decay as $t^{-12/5}$.  The dimensional argument
that captures these algebraic laws proceeds as follows.  The energy
decay is obtained by the energy balance $dE(t)/dt =-\epsilon(t)$,
where we estimate $E \sim v_{rms}^2(t)$ and 
$\epsilon(t) \sim v_{rms}^3(t)/L_{0}\sim E^{3/2}(t)/L_{0}$, and obtain 
$E(t) \sim t^{-2}$ and $\epsilon(t)\sim t^{-3}$.  As for the
vorticity decay, we have $\Omega(t) \sim (\delta_{\eta} v)^2/\eta^2$,
where $\eta$ is the dissipative lengthscale and $\delta_{\eta} v$ is
the typical velocity difference at separation $\eta$. Assuming a
Kolmogorov scaling $\delta_{\eta} v \sim \epsilon(t)^{1/3}\eta^{1/3}$, and 
recalling that for a second-order hyperviscous
dissipation $\epsilon(t)\sim \nu (\delta_{\eta} v)^2/\eta^{4}$ we
obtain $\eta \sim t^{3/10}$ and $\delta_{\eta} v \sim t^{-9/10}$,
whence $\Omega(t) \sim t^{-12/5}$.  We now focus on the process of
recovery of isotropy in terms of global quantities.  We identify two
set of observables 
\bea \Delta_{ij} E(t) &=& \la {E_i(t)-E_j(t)} \over{E_i(t) + E_j(t)} \ra\,,\\ 
\Delta_{ij} {\Omega}(t) &=& \la { \Omega_i(t)-\Omega_j(t)} \over {\Omega_i(t)
  + \Omega_j(t)} \ra\,, \eea 
which vanish for isotropic statistics.  Their rate of decay is therefore 
a direct measurement of the return to isotropy. The energy matrix $E_{ij}$ 
is particularly sensitive to the large scales while small-scale
fluctuations are sampled by $\Omega_{ij}$.  As it can be seen from
Fig.~\ref{fig:3}, both large and small scales begin to isotropize
after roughly one eddy turnover time and become fully isotropic
(within statistical fluctuations) after $100$ eddy turnover
times. However, small scales show an overall degree of anisotropy much
smaller than the large scales.
%----------------------------------------------------
%-----------------------------------------------
%---------------------------------------------------------
\section{The decay of small-scale fluctuations}
\label{longtimes}
\subsection{The self-preservation hypothesis for anisotropic decay}
The observables which characterize the decay of small-scale velocity
fluctuations are the longitudinal structure functions 
\be 
S^{(n)}\lp\r,t\rp = \la\left[\lp \v\lp\x+ \r,t\rp-\v\lp\x,t\rp\rp
  \cdot\hat\r\right]^n \ra\,.
\label{eq:longitudinal} 
\ee 
Those quantities depend not only on the modulus of the separation
$\r$ but also on its orientation $\hat\r$, since the velocity
statistics is anisotropic.  A method to systematically disentangle
isotropic from anisotropic contributions in the structure functions is
based on the irreducible representations of the SO(3) group
\cite{alp00}. In this approach, the observables
(\ref{eq:longitudinal}) are expanded on the complete basis of the
eigenfunctions of the rotation operator. The SO(3) decomposition of
scalar objects, such as structure functions, is obtained by projection
on the spherical harmonics $Y_{jm}(\unitr)$: 
\be 
S^{(n)}(\r,t) =\sum_{j=0}^{\infty}\sum_{m=-j}^{j} 
S^{(n)}_{jm}\lp r,t\rp Y_{jm}(\unitr).
\label{so3_sf} 
\ee 
Here, $S^{(n)}_{jm}\lp r,t\rp$ denotes the projection of the
$n$-th order structure function on the $(j,m)$ SO(3) sector, with $j$
and $m$ labeling the total angular momentum and its projection 
in the direction
$\unitz$, respectively.  Another equivalent possibility is to look at
the SO(3) decomposition of the PDF of the longitudinal velocity
differences. In this case, denoting by $\cP(\Delta,\r;t)$ the
probability that the longitudinal incremement $\dv \equiv \lp\v(\r,t)-\v(0,t)
\rp \cdot \unitr$ be equal to $\Delta$, we may project $\cP(\Delta,\r;t)$ on
the SO(3) basis: 
\be \cP(\Delta,\r;t) =\sum_{j=0}^{\infty}\sum_{m=-j}^{j} 
\cP_{jm}\lp r,\Delta;t\rp Y_{jm}(\unitr).
\label{so3_pdf} 
\ee 
The projection $\cP_{jm}\lp r,\Delta;t\rp$ plays the role of an
``effective PDF'' for each single SO(3) sector.  (It should be however
remarked that only the isotropic probability density $\cP_{00}\lp
r,\Delta;t\rp$ 
has the property of being everywhere positive and normalized to 
unity with respect to the weight $r^2/(4\pi)$.) Indeed, the
projections of the longitudinal structure function on any sector
$(j,m)$ can be reconstructed from the corresponding $\cP_{jm}\lp
r,\Delta;t\rp$ by averaging over all possible $\Delta$'s: 
\be
S^{(n)}_{jm}\lp r,t\rp = 
\int d\Delta \Delta^n \cP_{jm}\lp r,\Delta;t\rp\,.
\label{pdf_jm} 
\ee 
That establishes the equivalence between the decompositions
(\ref{so3_sf}) and (\ref{so3_pdf}).\\ The main points broached here
are about the long-time properties of the SO(3) projections,
$S^{(n)}_{jm}\lp r,t\rp$.  A simple isotropic generalization of the
self-preservation hypothesis (see, e.g.~Ref.~\cite{frisch}) amounts to
writing:
\be S^{(n)}_{jm}\lp r,t\rp = 
V^{(n)}_{jm}\lp t \rp f^{(n)}_{jm} \lp r/L_{jm}(t) \rp .
\label{eq:sph_jm} 
\ee 
Note that $V^{(n)}_{jm}\lp t \rp$ takes explicitly into account
the fact that large-scale properties may depend in a nontrivial way
on both $(j,m)$ and the order $n$.  Furthermore, $L_{jm}(t)$ accounts
for the possibility that the characteristic length scale depend on the
SO(3) sector.  \\ In analogy with the observations made in the
stationary case \cite{gw98,arad_exp,abmp99,ks,bt00,bdlt02,war00,bv00} we
postulate a scaling behavior 
\be S^{(n)}_{jm}\lp r,t \rp \sim a^{(n)}_{jm}(t) \lp \frac{r}
{L_{jm}(t)}\rp^{\zeta_{j}^{(n)}}\,.
\label{anyt}
\ee 
The time behavior is encoded in both the decay of the overall
intensity, accounted by the prefactors $a^{(n)}_{jm}(t)$, and the
variation of the integral scales $L_{jm}(t)$. The representation
(\ref{anyt}) is the simplest one fitting the initial time statistics
for $t=0$ and agreeing with the evolution given by the self
preservation hypothesis in the isotropic case. The power law behavior
for $f^{(n)}_{jm} \lp r/L_{jm}(t) \rp$ can be expected only in a
time-dependent inertial range of scales $ \eta(t) \ll r \ll L(t)$.  As
for the exponents appearing in (\ref{anyt}), their values are
expectedly the same as in the stationary case.  In the latter
situation it has been shown that they are organized hierarchically
according to their angular sector $j$ \cite{bdlt02}\,: 
\be \zeta_{0}^{(n)} \le \zeta_{1}^{(n)} \le \cdots \zeta_{j}^{(n)} \le \cdots .
\label{gerazetap}
\ee
Since the isotropic sector has the smallest exponent, at any given time
and for a given intensity of the fluctuation (selected by the value of $n$)
we have a recovery of isotropy going to smaller and smaller scales.
Yet, deviations of
the scaling exponents from their dimensional expectations make the recovery
at small-scales much slower than what predicted by
dimensional analysis \cite{bv00,war00}. Moreover, there are
quantities which should vanish in an isotropic field and actually
blow up as the scale decreases \cite{war00}.\\
Concerning the time evolution, it seems difficult 
to disentangle the dependence due to the 
decay of $a^{(n)}_{jm}(t)$
from the one due to the growth of the integral scale $L_{jm}(t)$.
Here, we note only that the
existence of a running reference scale, $L_{jm}(t)$ introduces some
non-trivial relations between the spatial anomalous scaling and the
decaying time properties, and those relations  
might be subject to experimental verification. 
In our case, the fact that the initial condition has a characteristic 
lengthscale comparable with the box size, simplifies the matter.
Indeed we expect that $L_{jm}(t) \approx L_{0}$, and the decay is due only 
to the fall off of the global intensity  $a^{(n)}_{jm}(t)$.
Unfortunately, a shortcoming is that the width of the inertial range
$L_{0}/\eta(t)$ shrinks monotonically in time, thereby limiting
the possibility of precise quantitative statements.
%%%%%%%%%%%%%%%%%%%%%%%%%%%%%%%%%
\subsection{Numerical results}
\begin{figure}[h] 
\epsfxsize=7.9truecm
\epsfysize=6.0 cm
\epsfbox{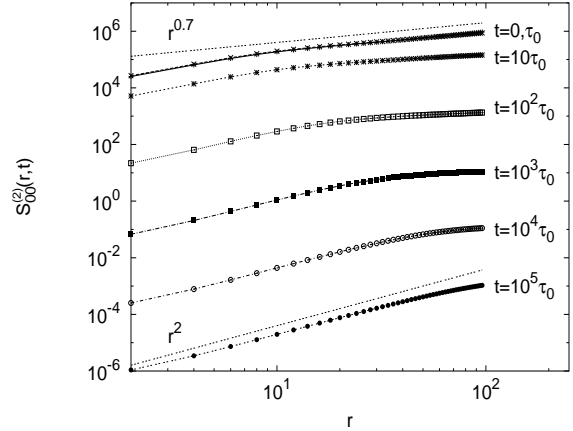}
\caption{Log-log plot of the isotropic component of the second order 
projection $S^{(2)}_{00}\lp r,t \rp $ {\it vs.} $r$ for $7$ decay 
times, $t=0,\tau_0,10\tau_0,10^2\tau_0,10^3\tau_0,10^4\tau_0,10^5\tau_0$ (from 
top to bottom).  The two straight lines correspond to the inertial 
range slope $S^{(2)}_{00}\lp r,t \rp \sim r^{0.7}$ (top) and to the 
smooth differentiable slope $S^{(2)}_{00}\lp r,t \rp \sim r^2$ (bottom).}
\label{fig:4} 
\end{figure}

\begin{figure}[h] 
\epsfxsize=7.9truecm
\epsfysize=6.0 cm
\epsfbox{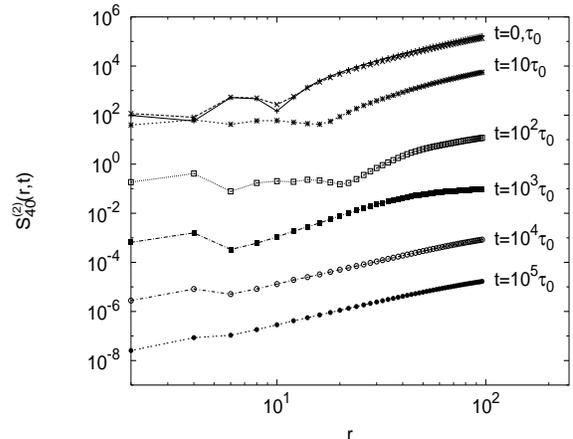}
\caption{The same quantities as in the previous figure but for the 
anisotropic sector $j=4,m=0$, i.e. log-log plot of $S^{(2)}_{40}\lp r,t \rp$ 
{\it vs.} $r$.}
\label{fig:5} 
\end{figure}

\begin{figure}[h] 
\epsfxsize=8.0truecm
\epsfysize=6.0 cm
\epsfbox{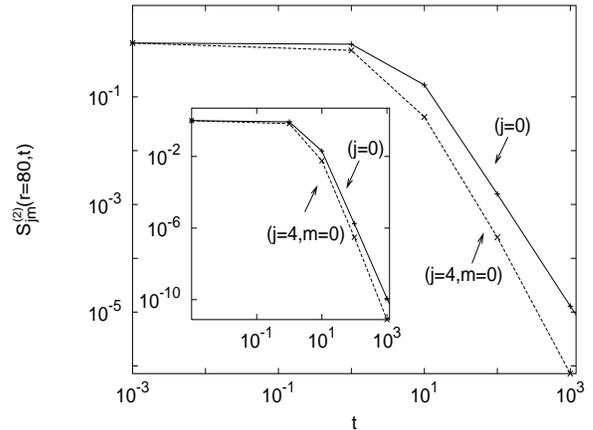}
\caption{Log-log plot of the second-order moment $S^{(2)}_{jm}\lp r=80,t \rp$
  for the isotropic sector $(j=0,m=0)$ and the anisotropic sector 
$(j=4,m=0)$, {\it vs.} time, for a separation $r$ in the inertial range. In 
the inset we plot the same curves for the fourth-order moment 
$S^{(4)}_{jm}\lp r=80,t\rp$.}
\label{fig:8} 
\end{figure}

\begin{figure}[h] 
\epsfxsize=7.9truecm
\epsfysize=6.0 cm
\epsfbox{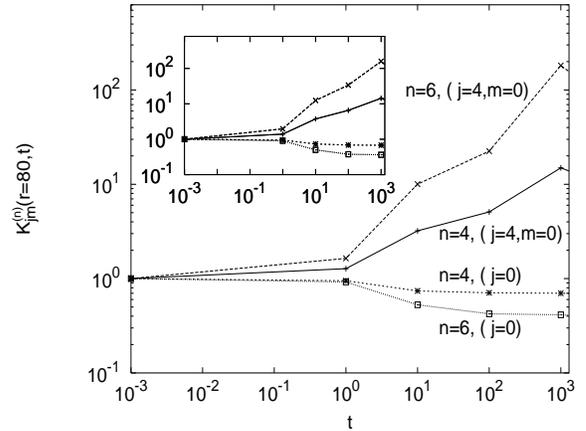}
\caption{Log-log plot of the generalized flatness, $K_{jm}^{(n)}(r,t)$
of order $n=4,6$ for both the isotropic (two bottom curves), and 
the anisotropic sector $(j=4,m=0)$ (two top curves) at $r=80$, and as
a function of time. In the inset we plot the same quantities, in the same
order, at a different inertial range scale, $r=40$.}
\label{fig:10} 
\end{figure}

An overall view of the SO(3) projections at all resolved scales and for
all measured decay times is presented in Fig.~\ref{fig:4} for
$n=2$ and the isotropic, $(j=0,m=0)$, sector.  The same quantities are
presented in Fig.~\ref{fig:5} for the most intense
anisotropic sector $(j=4,m=0)$. 
%---------------------------------------------------------
%-----------------------------------------------------------------
%-----------------------------------------------------------------
We notice that as time elapses the dissipative range erodes the inertial one,
as a consequence of the growth of the Kolmogorov scale $\eta(t)$.
We notice in passing that the two curves corresponding to $t=0$ and
$t=\tau_0$ almost coincide, i.e. even small scales are unchanged
despite their typical eddy turnover times is much smaller than
$\tau_0$. This finding has some consequences that will be discussed 
at length in section \ref{shorttimes}.  
A similar qualitative trend is displayed by the most
intense anisotropic sector, $(j=4,m=0)$ shown in Fig.~\ref{fig:5},
even though oscillations at small scales spoil significantly 
the scaling properties at small separations $r$.
%--------------------------------------------------------------
\begin{figure}[h] 
\epsfxsize=7.9truecm
\epsfysize=6.0 cm
\epsfbox{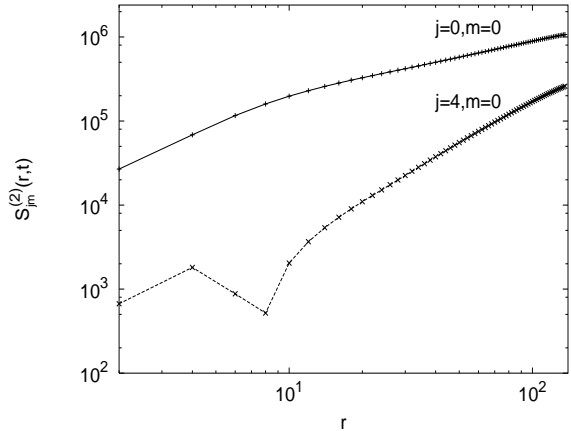}
\caption{Log-log plot of $S^{(2)}_{jm}\lp r,t \rp $ {\it vs.} $r$, at 
$t=\tau_0$. 
Symbols refer to $(j=0,m=0)$ (top) and to $(j=4,m=0)$ (bottom).}
\label{fig:6} 
\end{figure} 
%---------------------------------------------------------------
\begin{figure}[h] 
\epsfxsize=7.9truecm
\epsfysize=6.0 cm
\epsfbox{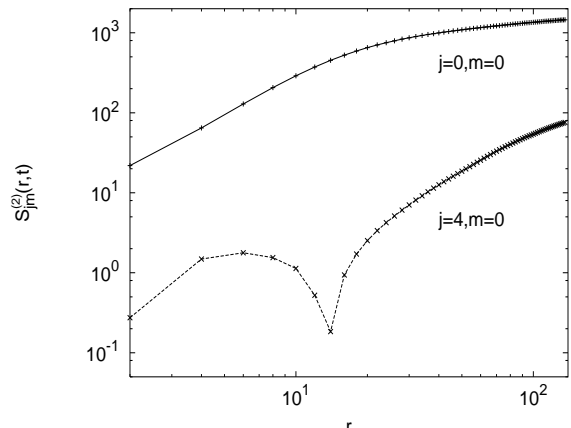}
\caption{The same quantities as in the previous figure but at the
later time $t=10^2 \tau_0$. The sectors shown are $(j=0,m=0)$ and $(j=4,m=0)$ 
(top to bottom), as in the previous figure.}
\label{fig:7} 
\end{figure} 
%----------------------------------------------------------------
In order to assess the relative importance of isotropic and
anisotropic contributions, we plot in Fig.~\ref{fig:6} the
SO(3) projections of some $(j,m)$ sectors for the order $n=2$
and a fixed time, $t=\tau_0$. Fig.~\ref{fig:7} shows the
same quantities as in Fig.~\ref{fig:6} but at a later time,
$t=100\tau_0$. Although the small-scale
behavior of anisotropic sectors readily becomes rather noisy, the various
contributions are organized hierarchically and the isotropic contribution is
dominant, as expected.\\
Let us now analyze quantitatively the time-decay of the structure functions
at a fixed separation. 
In Fig.~\ref{fig:8} we show the long-time
decay of the second and fourth-order moments on the
isotropic and an anisotropic sector at $r=80$, within the inertial range. 
%---------------------------------------------------------------------
%----------------------------------------------------------------
%-----------------------------------------------------------
\begin{figure}[h] 
\epsfxsize=8.0truecm
\epsfysize=6.0 cm
\epsfbox{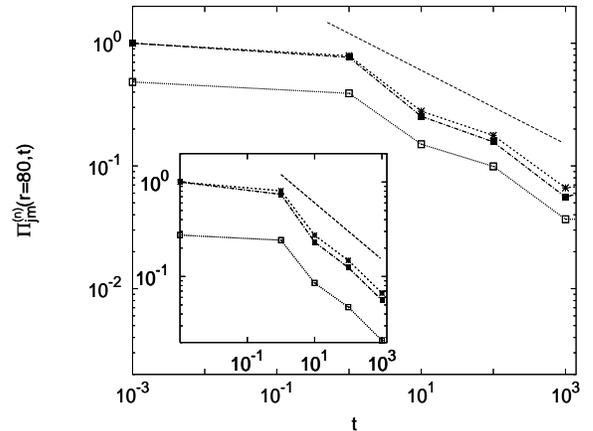}
\caption{Hierarchical organization of anisotropic fluctuations at long
times. Log-log plot of the anisotropic projections normalized by the
corresponding isotropic projection (see text), at two fixed scales $r=80$ and
$r=40$ (inset) for $n=2, 4, 6$ in the anisotropic sector $j=4,m=0$. 
Symbols read as follows\,: $\Pi^{(2)}_{40}$ (full box); 
$\Pi^{(4)}_{40}$ (star);
 $\Pi^{(6)}_{40}$ (empty box). The straight line 
is  $t^{-\chi^*}$ with $\chi^*\sim 0.3$.
 Same symbols in the inset.}
\label{fig:9} 
\end{figure} 
%-------------------------------------------------------------------
We observe that the anisotropic sectors decay faster than the
isotropic one, i.e. at a fixed scale there is a tendency towards
the recovery of isotropy at large times.
The relative rate of decay can be quantified by the observable
\be \Pi^{(n)}_{jm}\lp r,t \rp \equiv
\frac{S^{(n)}_{jm}\lp r,t \rp}{S^{(n)}_{0,0}\lp r,t \rp} \sim
t^{-\chi_{j}^{(n)}}\,.
\label{ratio_n}
\ee
In Fig.~\ref{fig:9} we show $\Pi^{(n)}_{jm}\lp r,t \rp$ at $r=80$ for
structure functions of order $n=2,4,6$ and for $(4,0)$, one of the 
most intense anisotropic sectors. In the inset we
also plot the same quantities at the smaller scale $r=40$. All
anisotropic sectors, for all measured structure functions, decay
faster than the isotropic one. The measured slope in the decay is
about $\chi_j^{(n)} \sim 0.3$ almost independent, within the
statistical errors, of the order $n$. 
Note that these results agree with the simple picture that
the time-dependence in (\ref{anyt}) is entirely carried by the
prefactors $a_{jm}^{(n)}(t)$ and the value of the integral scales
$L_{jm}(t)$ is saturated at the size of the box. Indeed, by assuming
that large-scale fluctuations are almost Gaussian we have that the
leading time-dependency of $a_{jm}^{(2n)}$ is given by $a_{jm}^{(2)}
a_{00}^{(2n-2)}$. 
For the isotropic sector, $a_{00}^{(2n)} \sim (a_{00}^{(2)})^n$, and plugging
that 
in (\ref{ratio_n}), we get: 
$\Pi^{(n)}_{jm}\lp r,t \rp \sim a^{(2)}_{jm}(t)/a^{(2)}_{00}(t) \sim
t^{-\chi^*}$ 
with $\chi^* \sim 0.3 (\pm 0.1)$ independent of $n$. The quality of our data
is insufficient to detect possible residual effects due to $L_{jm}(t)$, which
would make $\chi_j^{(n)}$ depend on $n$ and $j$ because of spatial 
intermittency.\\
The interesting fact that we measure decay properties of the
anisotropic sectors which are almost independent of the order of the
structure functions indicate that we must expect some non-trivial time
dependence in the shape of the PDF's $\cP_{jm}\lp r ,\Delta;t\rp$ for
$j>0$. The most accurate way to probe the rescaling properties of
$\cP_{jm}\lp r,\Delta;t\rp$ in time is to compute the generalized
flatness: 
\be 
K_{jm}^{(n)}(r,t)\equiv \frac{S^{(n)}_{jm}\lp r,t \rp}{\lp S^{(2)}_{jm}\lp r,t
  \rp
 \rp^{\frac{n}{2}}} \sim t^{\alpha_{j}^{(n)}} 
\ee 
Were the PDF projection in the $(j,m)$ sector 
self-similar for $t \gg \tau_0$,
then $K_{jm}^{(n)}(r,t)$ would tend to costant
values. This is not the case for anisotropic fluctuations,
as it is shown in Fig. (\ref{fig:10}). The curves
$K_{jm}^{(n)}(r,t)$ are collected for two fixed inertial range
separations, $r=80$ and $r=40$ (inset), for two different orders,
$n=4,6$ and for both the isotropic and one of the most intense 
anisotropic sectors$(j=4,m=0)$ . The isotropic flatness tends toward a constant
value for large $t$. Conversely, its anisotropic counterparts are
monotonically increasing with $t$, indicating a tendency for the
anisotropic fluctuations to become more and more intermittent as time
elapses. A consequence of the monotonic increase of intermittency for large
times is the impossibility to find a rescaling function, $g(t,r)$,
which makes the rescaled PDF $g(t,r)\cP_{jm}\lp r,\frac{\Delta}{g(t,r)};t\rp$ 
time-independent at large times. Let us notice that also the behavior in
Fig.~\ref{fig:10} is in qualitative agreement with
the observation previously made that all time dependencies can be
accounted by the prefactors $a_{jm}^{(n)}(t)$. Indeed, assuming that the
length scales $L_{jm}(t)$ have saturated and that the large scale PDF is 
close to Gaussian, it is easy to work out the prediction
$K_{jm}^{(n)}(r,t) \sim t^{-\chi^*(1-n/2)}$, i.e.
$\alpha_{j}^{(n)}=\chi^*(n/2-1)$.
%------------------------------------------------------------------
%-------------------------------------------------------------
 We conclude this
section by a brief summary of the results.  We have found that
isotropic fluctuations persist longer than anisotropic ones,
i.e. there is a time-recovery, albeit slower than predicted by dimensional
arguments, of isotropy during the decay process.  We have also found
that isotropic fluctuations decay in an almost self-similar way whilst
the anisotropic ones become more and more intermittent. Qualitatively,
velocity configurations get more isotropic but anisotropic
fluctuations become, in relative terms, more ``spiky'' than the
isotropic ones as time elapses.
%%%%%%%%%%%%%%%%%%%%%%%%%%%%%%%%%%%%%%%%%%%%
\section{Short-time decay} 
\label{shorttimes} 
\begin{figure}[h] 
\epsfxsize=8.3truecm
\epsfysize=6.0 cm
\epsfbox{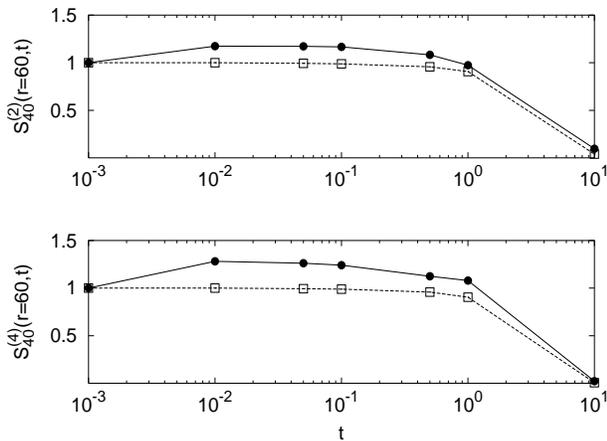}
\caption{The same curves as in the previous figure but for the
anisotropic sector $(j=4,m=0)$.}
\label{fig:12} 
\end{figure}

Let us now move to the properties of decay at short times ($t \ll \tau_0$). 
Universality of small-scale forced turbulence is at
the forefront of both theoretical and experimental investigation of
real turbulent flows \cite{frisch}. The problem is to identify those
statistical properties which are robust against changes of the
large-scale physics, that is against changes in the boundary
conditions and the forcing mechanisms. Our goal here is to relate the
small-scale universal properties of forced turbulent statistics
to those of short-time decay for an ensemble of initial
configurations. An immediate remark is that one cannot expect
an universal behaviour for all statistical observables
as the very existence of anomalous scaling is the
signature of the memory of the boundaries and/or the external forcing
throughout all the scales. 
Indeed, the main message we want to convey here is that only 
the scaling of both isotropic and anisotropic small-scale fluctuations 
is universal, at least for forcings concentrated at large scales. 
The prefactors are not expected to be so. 
There is therefore no reason to expect that
quantities such as the skewness, the kurtosis and in fact the whole
PDF of velocity increments or gradients be universal. \\ This is the
same behavior as for the passive transport of scalar and vector fields
(see \cite{rev} and references therein). For those systems both the
existence and the origin of the observed anomalous scaling laws have
been understood and even calculated analytically for some instances in
the special class of Kraichnan flows \cite{k94}. Here, it is worth
stressing that the universal character of scaling exponents is shared
by both isotropic and anisotropic fluctuations \cite{isoani}. \\ For
the Navier-Stokes case we have a huge amount of experimental and
numerical indications that the velocity field shows anomalous scaling
\cite{frisch}, suggesting the existence of phenomena ``similar'' to
those of the linear case. However, carrying over the analytical
knowledge developed for linear hydrodynamical problems involve some
nontrivial, yet missing, steps. For the Navier-Stokes dynamics, linear
equations of motion surface again but at the functional level of the
whole set of correlation functions. In a schematic form: 
\be
\partial_t C^{(n)} = \Gamma^{(n+1)}C^{(n+1)} +\nu D^{(n)} C^{(n)} + F^{(n)},
\label{compact} 
\ee 
where $\Gamma^{(n+1)}$ is the integro-differential linear
operator coming from the inertial and pressure terms and $C^{(n+1)}$
is a shorthand notation for a generic $(n+1)$-point correlator.  The
molecular viscosity is denoted by $\nu$ and $D^{(n)}$ is the linear
operator describing dissipative effects. Finally, $F^{(n)}$ is the
correlator involving increments of the large-scale forcing $\f$ and of
the velocity field.
Balancing inertial and injection terms gives dimensional scaling, 
and anomalously scaling terms must therefore have a different source. 
A natural possibility is that a mechanism similar to the one identified
in linear transport problems be at work in the Navier-Stokes case as
well. The anomalous contributions to the correlators would then be
associated to statistically stationary solutions of the unforced
equations (\ref{compact}). The scaling exponents would {\it a fortiori} be
independent of the forcing and thus universal. As for the prefactors,
the anomalous scaling exponents are positive and thus the anomalous
contributions grow at infinity. They should then be matched at the
large scales with the contributions coming from the forcing to ensure
that the resulting combination vanish at infinity, as required for
correlation functions. Our aim here is not to prove the previous
points but rather to check over the most obvious catch: the
Navier-Stokes equations being integro-differential, non-local
contributions might directly couple inertial and injection scales and
spoil the argument. This effect might be particularly relevant for
anisotropic fluctuations where infrared divergences may appear in the
pressure integrals \cite{ap01}. 
%----------------------------------------------------------- 
\begin{figure}[h] 
\epsfxsize=8.2truecm
\epsfysize=6.0 cm
\epsfbox{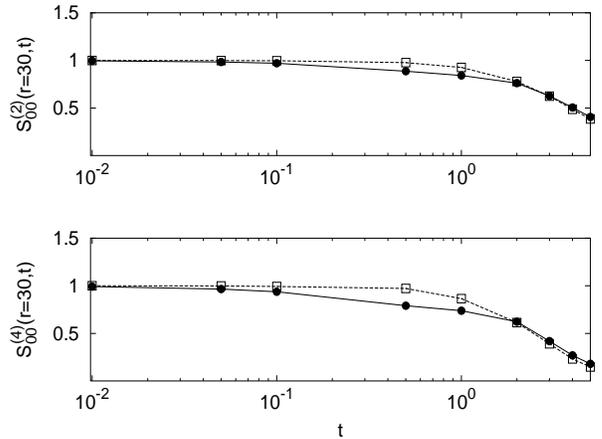} 
\caption{Top: Temporal decay of the second-order isotropic structure
function $S^{(2)}_{00}(r,t)$, rescaled by its value at $t=0$. 
Here $r=30$, inside the inertial
range. The two curves refer to the time evolution of the structure
function starting from the forced-stationary velocity
fields (squares, set A) and from the randomly dephased velocity
fields (circles, set B). Time is normalized by the integral eddy turnover
time.
Notice that for set B we observe changes on a time
scale faster than the integral eddy turnover time. That is to be
contrasted with the case A, where structure functions are strictly constant
in time up to an integral eddy turnover
time. Bottom: The same curves but for the fourth-order structure function.}
\label{fig:11} 
\end{figure} 
%-----------------------------------------------------------------------
In order to investigate the previous point, we performed two sets of
numerical experiments in decay.  The first set, A, is of the same kind
as in the previous section, i.e. we integrated the unforced
Navier-Stokes equations (\ref{eq:navierstokes})
with initial conditions picked from an ensemble obtained from a forced
anisotropic stationary run.  Statistical observables are measured as
an {\it ensemble} average over the different initial conditions,
$\langle \bullet \rangle_{ens}$. The ensemble at the initial time of
the decay process is therefore coinciding with that at the stationary
state in forced runs. If correlation functions are indeed dominated
at small scales by statistically stationary solutions of the unforced
equations then the field should not decay. Specifically, the field
should not vary for times smaller than the large-scale eddy turnover
time $\tau_0 \sim L_0/\langle v^2 \rangle^{1/2}$, with $L_0$ denoting
the integral scale of the flow. Those are the times when the effects
of the forcing terms start to be felt. Note that this should hold
at all scales, including the small ones whose turnover times are much
faster than $\tau_0$. The second set of numerical simulations (set B)
is meant to provide for a stringent test of comparison. The initial
conditions are the same as before but for the random scrambling of the
phases\,: $\hat{v}_i(\k) \rightarrow P_{il}(\k)\,
\hat{v}_l(\k)\,\exp(i\,\theta_l(\k))$. 
Here, $\hat{v}$ denotes the Fourier transform and $P_{il}(\k)$ is the 
incompressibility projector. In this way, the spectrum and its scaling 
are preserved but the wrong organization of the phases is expected to spoil the
statistical stationarity of the initial ensemble. As a consequence,
two different decays are expected for the two sets of experiments. In
particular, contrary to set A, set B should vary at small scales on
times of the order of the eddy turnover times $\tau_r \sim r^{2/3}$.
This is exactly what we found in the numerical simulations, as can be
seen in Fig.~\ref{fig:11}, where the temporal behavior of
longitudinal structure functions of order 2 and 4 is shown. The
scaling of the contributions responsible for the observed behaviour at
small scales are thus forcing independent.\\
As for anisotropic fluctuations, we also found two very different
behaviors depending on the set of initial conditions.  In
Fig.~\ref{fig:12} we show the case of the projection
$S^{(n)}_{jm}(r=60,t)$\, for the anisotropic sector $j=4$,$m=0$. As it can
be seen, for set A of initial conditions the function is indeed not
decaying up to a time of the order of $\tau_0$.\\
To conclude, the data presented here support the conclusion that
nonlocal effects peculiar to the Navier-Stokes dynamics do not spoil
arguments on universality based on analogies with passive turbulent
transport. The picture of the anomalous contributions to the
correlation functions having universal scaling exponents and
non-universal prefactors follows.
%---------------------------------------------------------
%--------------------------------------------------------------- 
\section{Conclusions}
\label{conclusions}
We have presented a numerical investigation of the decay of
three-dimensional turbulence, both isotropic and anisotropic.
Concerning short-time decay, we have compared the decay for
two different sets of initial conditions, with and without phase
correlations.  That gave some new hints on the properties of
universality of isotropic and anisotropic forced turbulence.
As for long times, we have found that fluctuations in the
inertial range become more and more isotropic. On the other hand, the
anisotropic components become more and more intermittent,
i.e. relatively intense anisotropic fluctuations become more and more
probable. The main issue here was to investigate the effects of a
bounded domain, i.e.  situations where the integral scale cannot grow
indefinitely. The decay process at long times is then governed by the
set-up at large scales. Anisotropies decay in time at a rate almost
independent of the order of the moments and of the kind of anisotropic
fluctuations. Projections of the PDF's on different SO(3) sectors show
different intermittent properties (with the anisotropic sectors being
more intermittent). An obvious further development of this study would
be to investigate the case where the integral scales $L_{jm}(t)$ vary
in time. Additional intermittency in time might then be brought in by
the anomalous scaling in the space variables of the 
correlation functions.\\

This research was supported by the EU under the Grants
No. HPRN-CT 2000-00162 ``Non Ideal Turbulence'' and
No. HPRN-CT-2002-00300 "Fluid Mechanical Stirring
and Mixing", and by the INFM
(Iniziativa di Calcolo Parallelo). 
A.L. acknowledges the hospitality of the Observatoire de la C\^{o}te d'Azur, 
where part of the work has been done.
%%%%%%%%%%%%%%%%%%%%%%%%%%%%%%%%%% 
 
\end{multicols} 
\end{document}